\documentclass[review,3p]{elsarticle}
\usepackage{hyperref}
\usepackage{siunitx}
\usepackage{xspace}
\usepackage{xfrac}
\usepackage{amsmath}
\usepackage{multicol}
\usepackage{textcomp}
\usepackage{url}
\usepackage[english]{babel}
\usepackage{lineno}
\usepackage{xcolor}

\journal{Nuclear Materials \& Energy}
\begin{document}
\hypersetup{
pdftitle={Integrated modeling of boron powder injection for real-time plasma-facing component conditioning},
pdfsubject={plasma-material interactions},
pdfauthor={Florian Effenberg},
pdfkeywords={boronization, mixed-materials, walldyn3d, emc3-eirene, dust transport, impurity powder dropper}
}

\begin{frontmatter}
\title{Integrated modeling of boron powder injection for real-time plasma-facing component conditioning}

\author[pppl]{F. Effenberg\corref{correspondingauthor}}
\cortext[correspondingauthor]{Corresponding author}
\ead{feffenbe@pppl.gov}
\author[ipp1]{K. Schmid}
\author[pppl]{F. Nespoli}
\author[pppl]{A. Bortolon}
\author[ipp2]{Y. Feng}
\author[ga]{B.A. Grierson}
\author[ornl]{J.D. Lore}
\author[pppl]{R. Maingi}
\author[ucsd]{D.L. Rudakov}

\address[pppl]{Princeton Plasma Physics Laboratory, Princeton, NJ 08543, USA}
\address[ipp1]{Max-Planck-Institut f\"ur Plasmaphysik, 85748 Garching b. M\"unchen, Germany}
\address[ipp2]{Max-Planck-Institut f\"ur Plasmaphysik, 17491 Greifswald, Germany}
\address[ga]{General Atomics, San Diego, CA 92186, USA}
\address[ornl]{Oak Ridge National Laboratory, Oak Ridge, TN 37831, USA}
\address[ucsd]{University of California - San Diego, La Jolla, CA 92093, USA}

\begin{abstract}
An integrated modeling framework for investigating the application of solid boron (B) powder injection for real-time surface conditioning of plasma-facing components (PFCs) in tokamak environments is presented. Utilizing the DIII-D impurity powder dropper (IPD) setup, this study simulates B powder injection scenarios ranging from milligrams to tens of milligrams per second, corresponding to boron flux rates of $10^{20}-10^{21}$ B/s in standard L-mode conditions. The comprehensive modeling approach combines EMC3-EIRENE for simulating the deuterium plasma background and the Dust Injection Simulator (DIS) for the ablation and transport of the boron powder particles. EMC3 trace impurity fluid modeling results show substantial boron transport to the inboard lower divertor, predominantly influenced by the main ion plasma flow. The dependency on powder particle size (5-250 $\mu$m) was found to be insignificant for the scenario considered. The effects of erosion and redeposition were considered to reconcile the discrepancies with experimental observations, which saw substantial deposition on the outer divertor plasma-facing components. For this purpose, the WallDYN3D code was updated to include boron sources within the plasma domain and integrated into the modeling framework. The mixed-material migration modeling shows evolving boron deposition patterns, suggesting the formation of mixed B-C layers or predominantly B coverage depending on the powder mass flow rate. While the modeling outcomes at lower B injection rates tend to align with DIII-D experimental observations, the prediction of near-pure boron layers at higher rates has yet to be experimentally verified in the carbon environment of the DIII-D tokamak. The extensive reach of boron layers found in the modeling suggests the need for modeling that encompasses the entire wall geometry for more accurate experimental correlations. This integrated approach sets a precedent for analyzing and applying real-time in-situ boron coating techniques in advanced tokamak scenarios, potentially extendable to ITER.
\end{abstract}

\end{frontmatter}
 
\section{Introduction\label{sec:Intro}}
Real-time wall conditioning (RTWC) has recently gained attention in tokamak and stellarator studies~\cite{bortolon_real-time_2019, bortolon_observations_2020, lunsford_characterization_2021, gilson_wall_2021, krieger_2022, effenberg_2022, bodner_2022, xu_2023, effenberg_2023}. Traditional techniques, like glow discharge boronization, use hazardous borane gases and interrupt plasma operation, making them unsuitable for long-pulse reactors. They also condition divertor PFCs less effectively than the main chamber walls. For ITER, the strategic decision to replace the beryllium wall with a full tungsten (W) one offers benefits such as better tritium compatibility, lower dust production, and enhanced resilience to thermal loads during disruptions. While current analysis confirms that Q=10 operation is achievable, investigating additional risk-mitigation strategies, such as boron material injection, remains beneficial \cite{loarte_2024}. Similarly, in the context of the W7-X and DIII-D wall change-out projects \cite{fellinger_2023, abrams_2024}, which are considering tungsten as a reactor-relevant material, real-time boronization may facilitate high-Z impurity control and support access to low-collisionality plasmas. The development of impurity powder dropper (IPD) technology \cite{nagy_multi-species_2018}, which releases various impurity powders into the plasma in real time, has significantly advanced RTWC research. The IPD dispenses non-gaseous impurities like lithium, boron, boron nitride, and silicon. Understanding the transport and deposition of these impurities in the scrape-off layer (SOL) is crucial for ITER and future fusion devices.

This paper presents a new framework for modeling solid boron injection, the key material for real-time wall conditioning in tokamaks, stellarators, and heliotrons. It integrates transport, erosion, and deposition modeling to assess boron coverage on divertor PFCs, crucial for evaluating its suitability in future devices like ITER and Fusion Pilot Plants (FPPs) \cite{wauters_2020, stangeby_2022, snipes_2024}. As a proof of concept, results from the new workflow, which couples three key models — EMC3-EIRENE, DIS, and WallDYN3D — based on a DIII-D RTWC scenario from a previous study \cite{effenberg_2020}, will be evaluated.

The EMC3-EIRENE code \cite{feng_recent_2014, reiter_2005} simulates full torus plasma, impurity transport, and plasma-material interactions. It has revealed non-axisymmetric effects from atomic impurity sources in axis-symmetric setups, such as nitrogen seeding in Alcator C-Mod and boron injection in DIII-D \cite{lore_three-dimensional_2015, effenberg_2020}. The Dust Injection Simulator (DIS) \cite{nespoli_2021} models dust transport, similar to DUSTT, which has been applied to intrinsic dust in 2D plasma \cite{smirnov_2007, smirnov_2023} and injected dust in the 3D heliotron LHD \cite{shoji_2020}.

To decipher the mechanisms that lead to B wall coatings through B powder injection, it is crucial to correlate deposition with transport. For this purpose, the WallDYN3D code was employed, which has been recently extended to 3D configurations to enable studies in both tokamaks with non-axisymmetric wall geometries, RMPs, or other symmetry breaking such as local gas puffs and inherently three-dimensional configurations like stellarators \cite{schmid_2011, schmid_2018, schmid_2020}.
The forthcoming sections of this paper briefly introduce the three models used (Sec. \ref{sec:EMC3DIS} and \ref{sec:WallDYN3D}), discuss the transport effects, and resulting B distribution without impurity recycling (Sec. \ref{sec:Transport}), proceed with the analysis based on mixed-material modeling (Sec. \ref{sec:Mixedmaterial} and \ref{sec:BCratio}), and finally, present a summary and conclusions (Sec. \ref{sec:Conclusion}).
\section{Coupled dust and plasma impurity transport modeling with DIS and EMC3-EIRENE \label{sec:EMC3DIS}}
For the present numerical study, the 3D plasma fluid and kinetic neutral edge transport (EMC3-EIRENE) have been coupled with the Dust Injection Simulator (DIS). EMC3 solves a set of reduced Braginskii fluid equations for particles, parallel momentum, and energies for electrons and ions \cite{feng_recent_2014}. EIRENE solves the kinetic transport equations for neutral atoms and molecules, including collisional processes \cite{reiter_2005}. EMC3-EIRENE can handle full-torus computational grids, including complex field and wall geometries.

The Dust Injection Simulator (DIS) model was designed to study the dynamics of dust particles within three-dimensional, time-evolving plasmas \cite{nespoli_2021}. It employs a comprehensive approach, integrating equations that account for the motion, charge balance, temperature changes, and mass evolution of the dust particles. These equations allow for the simulation of dust transport and ablation processes, including evaporation.

The equation of motion for dust particles of mass $m_d$, $m_d\frac{dv}{dt}=R_f$, can be described as such:
\begin{equation}
    R_f=F_{drag} + F_{E} + F_{grav} +F_{centr.}
\label{eq:imp_eq1}
\end{equation}
where $R_f$ describes the force acting on the powder particles consisting of the drag force exerted by the plasma ($F_{drag}$), electrostatic force ($F_{E}$), gravitational force ($F_{grav}$), and centrifugal force ($F_{centr.}$). Presently, $F_{E}$ does not contribute to the dust transport since EMC3-EIRENE does not yet solve the electric field self-consistently. Particles reaching the wall collide elastically, reflect, and continue evaporating in the plasma. The deposition depths in the plasma of the injected materials are determined by material-specific properties such as vapor pressure, heat sublimation, emissivity, secondary electron emission, material density, and boiling and melting points. Note that the present version of the model assumes perfect spherical symmetry of particles, neglecting effects such as the rocket force \cite{krasheninnikov_2010}. 

The powder migration and ablation generate a 3D distribution of atomic boron sources that contribute to a total boron influx (B/s), corresponding to the mass flow injection rate (mg/s). Once the neutral impurities are released and ionized, their transport is modeled through the trace impurity fluid module, which solves the following equations:
\begin{equation}
\nabla_{\parallel} \cdot (n_z V_{z\parallel}) + \nabla_{\perp} \cdot (-D_{imp}\nabla_{\perp}n_z) =S_z
\label{eq:imp_eq2}
\end{equation}
\begin{eqnarray}
0=-\frac{1}{n_z}\frac{dp_z}{ds}+ZeE_{\parallel}+m_z\frac{V_{\parallel}-V_{z\parallel}}{\tau_{zi}} \nonumber \\
+0.71Z^2\frac{dT_e}{ds}+2.6Z^2\frac{dT_i}{ds} \\
 =F_{P} + F_{E} + F_{fr} + F_{th,e} + F_{th,i}    \nonumber
\label{eq:imp_eq3}
\end{eqnarray}
where $S_z$ is the impurity ionization source for charge state $Z$, $D_{imp}$ the anomalous diffusivity, $n_z$ and $p_z$ the impurity density and pressure, $E_{\parallel}$ the parallel electric field, and $V_{\parallel}$ and $V_{z\parallel}$ the main and impurity ion parallel velocities. The impurity-main ion collision time is given by $\tau_{zi}=\frac{1.47\cdot 10^{13}m_z(\sfrac{T_i^3}{m_i})^{0.5}}{(1 + \sfrac{m_i}{m_z})n_iZ^2 ln\Lambda}$ with $\ln\Lambda \approx 15$ \cite{stangeby_plasma_2000}. Equation 3 terms correspond to pressure ($F_P$), electric ($F_E$), friction ($F_{fr}$), and electron/ion thermal forces ($F_{th,e}$, $F_{th,i}$). The impurity temperature is assumed to be equal to the main ion temperature ($T_z = T_i$).

In many plasma edge scenarios, electrostatic and pressure forces can be neglected, reducing equation 3 to focus on friction and ion thermal forces independent of charge state $Z$. At steady state, this leads to a flow velocity balance \cite{feng_physics_2006, effenberg_2020}:
\begin{equation}
V_{z\parallel}\approx V_{\parallel} - \alpha\frac{dT_i}{ds} = V_{fr} - V_{i-th}
\label{eq:imp_eq4}
\end{equation}
The flows $V_{fr}$ and $V_{i-th}$ result from friction and ion thermal forces, respectively. A one-dimensional core transport model \cite{feng_recent_2014} simulates the ionization of impurities within the core plasma. Despite the axis-symmetric magnetic equilibrium, the trajectories of powder particles and the resulting impurity ion distributions are three-dimensional.
\section{Coupling impurity transport with erosion and deposition dynamics in WallDYN3D\label{sec:WallDYN3D}}
In the WallDYN3D model, the evolution of surface density for materials on the reactor walls, crucial for assessing material behavior under plasma exposure, is described by differential equations that account for erosion, deposition, and material mixing processes. For this study, the WallDYN3D model was extended to include impurity sources located within the plasma domain resulting from the boron powder particle evaporation. The B atomic sources provided by the EMC3-EIRENE/DIS calculations are incorporated into the WallDYN3D model as a charge state ($qi$) resolved B injection influx ($\Gamma_{ei,wj,qi}^{Inj}$) at each wall segment ($wj$), where $wj$ represents a discretized location on the reactor wall.

The total impurity influx of component $ei$ on a wall element $wk$, defined as the sum of contributions from reflected, eroded/sublimated, and injected fluxes, is expressed as:
\begin{align}
\Gamma_{ei,wj,qi}^{In} &= \sum_{wk}m_{ei,wk,wj,qi}(\Gamma_{ei,wk}^{Ero}+\Gamma_{ei,wk}^{Refl})+\Gamma_{ei,wj,qi}^{Inj}
&\\ \nonumber
\Gamma_{ei,wk}^{Refl}&=\sum_{qi}(1-R_{ei,wk,qi})\Gamma_{ei,wk,qi}^{In}
&\\ 
\Gamma_{ei,wk}^{Ero}&=\sum_{em}\sum_{qm}\Gamma_{em,wk,qm}^{In}Y_{ei,em,wk,qm} & \nonumber
\label{eq:imp_eq5}
\end{align}
where $m_{ei,wk,wj,qi}$ describes the charge state-resolved redistribution and migration matrix, and $R_{ei,wk,qi}$, and $Y_{ei,em,wk,qm}$, describe the reflection and erosion yields (including charge-exchange erosion), respectively.
The governing equation for the areal density (atoms/m\(^2\)) time evolution of wall surface element $\delta_{ei,wk}$ is defined as:
\begin{equation}
    \frac{d\delta_{ei,wk}}{dt} = \sum_{qi}\Gamma_{ei,wk,qi}^{In}-(\Gamma_{ei,wk}^{Ero}+\Gamma_{ei,wk}^{Refl})+\Gamma_{ei,wk}^{Bulk},
\label{eq:imp_eq6}
\end{equation}
where $\Gamma_{ei,wk}^{Bulk}$ represents a bulk material reservoir, that maintains the areal density of all wall elements.
\begin{figure}[ht]
\begin{center}
\includegraphics[width=85mm]{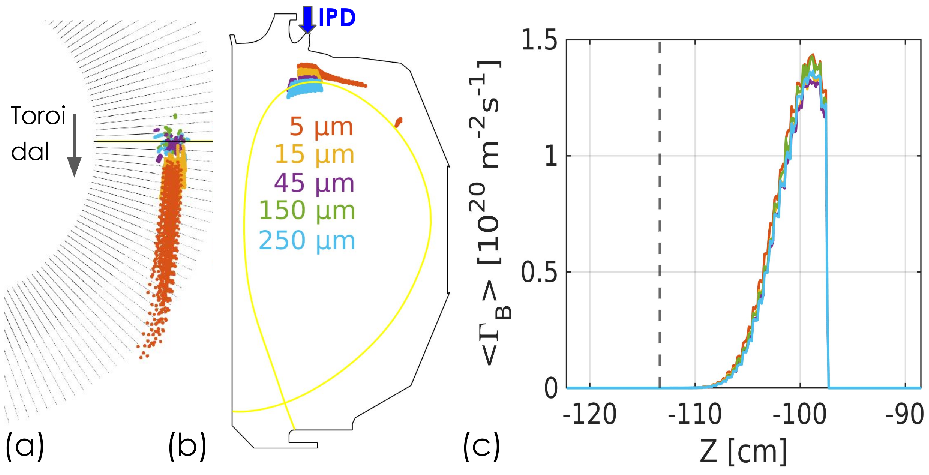}
\caption{\label{fig:figure1} (a) Top-down view and (b) cross-sectional view on boron source distribution resulting from ablation of 5-250 $\mu$m size powder particles along the trajectories calculated with DIS. (c) Boron flux profile at the lower inner divertor (peak location) for 5-250 $\mu$m B powder particles (zero impurity recycling), shown along the Z-coordinate of the inner divertor target. The vertical dashed line indicates the inner strike point position.}
\end{center}
\end{figure}
The redistribution matrix represents the probability that impurities, once eroded from a specific wall location, will migrate through the plasma and redeposit at a different wall location, potentially in a modified charge state. This description assumes that surface evolution occurs over a slower timescale ($\approx$s) compared to the faster timescales of impurity and plasma transport ($\approx$ms). A solution is achieved when the surface, including the thin Reaction Zone of about 1 nm, reaches a quasi-steady state where the net erosion and deposition rates balance. This stabilization of surface concentration indicates that the dynamic equilibrium of the surface under plasma interactions has been effectively captured.
Further details about the model can be found in \cite{schmid_2018,schmid_2020a,schmid_2020}.
\section{Full-torus modeling of boron powder injection and transport with EMC3-EIRENE and DIS\label{sec:Transport}}
A representative L-mode deuterium (D) plasma scenario for real-time wall conditioning in the DIII-D tokamak has been implemented using the EMC3-EIRENE code. This implementation utilizes a full-torus (360$^{\circ}$) grid based on the equilibrium $\#179894$ at 3500~ms, specifically designed to fully capture the migration of boron powder particles and the resulting distributions of boron sources and fluxes. Previously, this scenario was used to model boron transport from a single atomic source \cite{effenberg_2020}. In the current implementation, boron is introduced through powder injection, and the ablation and transport of this boron dust are simulated using the DIS code. A toroidal cross-section of the plasma grid domain for EMC3-EIRENE is shown in Figure \ref{fig:figure1}(b). The impurity powder source is situated at the tip of the upper small angle slot (SAS) divertor, where it is connected via a vertical tube to the powder dropper. The simulation accounts for a 5 cm diameter gap at this dropping location in the divertor geometry.

In the first step, the deuterium plasma and neutral transport have been calculated for the reference scenario without impurity injection for the entire torus. An upstream density of $n_{up}=1.5\cdot10^{19}$ m$^{-3}$ was set based on experimental measurements, resulting in recycling flux $\Gamma_{rec}=13.3$ kA. The anomalous particle and heat cross-field transport in the plasma edge was prescribed by $D_{\perp}=0.33$ m$^{2}$s$^{-1}$ and $\chi_{\perp}=1.0$ m$^{2}$s$^{-1}$ for the purpose of this study based on reference values \cite{effenberg_2020}. The cross-field diffusivity of main plasma and impurity ions is assumed to be equal for simplicity.

In the second step, the powder injection through the SAS port was modeled with DIS. A mass flow rate of $5$ mg/s was assumed, corresponding to an atomic rate of $2.8\times10^{20}$ B/s. Within the model, the gravitational force accelerates the powder particles downwards into the plasma, reaching particulate velocities of 5 m/s at the injection point. Random angular distribution is assumed to take deviations from perfect vertical trajectories due to collisions within the drop tube into account, which can lead to trajectories that have components in the radial and toroidal directions, slightly deviating from an ideal vertical trajectory. 
\begin{figure}[h!]
\begin{center}
\includegraphics[width=85mm]{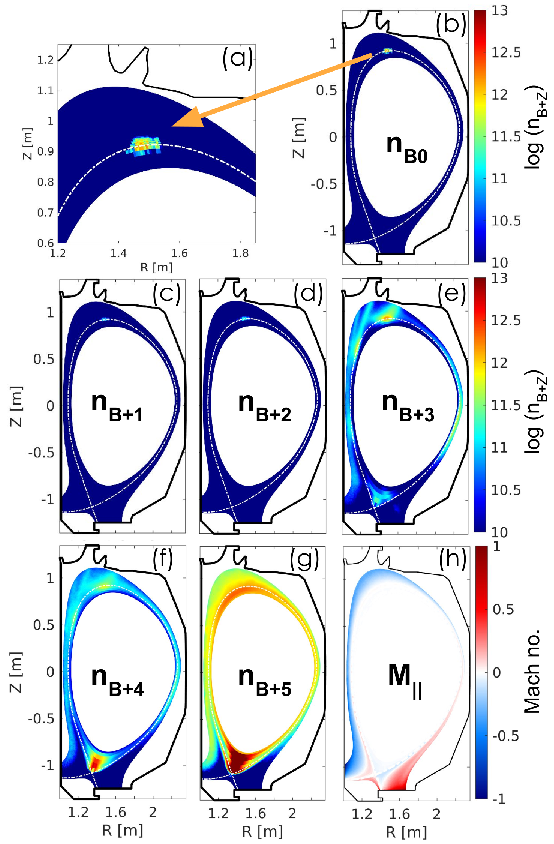}
\caption{\label{fig:figure2} (a, b) Distributions of neutral boron ($B^{0}$) and (c-g) ionized boron states ($B^{+1}$ to $B^{+5}$) at the powder injection location in the torus, considering 45 $\mu$m boron particles. (h) Parallel flow velocity profile (Mach number) with blue and red indicating flows toward the inner and outer divertor targets, respectively.}
\end{center}
\end{figure}
Figure \ref{fig:figure1}(a,b) displays boron source distributions along particle trajectories for sizes from 5 to 250 $\mu$m, representing typical experimental diameters. The injected mass flow rate is constant for all particle sizes. The evaporation usually occurs when the powder particles reach the separatrix, where the electron temperature and associated heat flux $Q_{evap}$ increase, raising the powder particle temperature to the evaporation point. For the mass loss rate follows: $\frac{dm_d}{dt}\propto Q_{evap}$. Larger particles, such as those measuring 150 and 250 $\mu$m, follow a predominantly vertical trajectory due to gravity and may cross the separatrix before fully ablating. Conversely, the trajectories of smaller particles (5 $\mu$m) are primarily influenced by the drag and centrifugal forces exerted by the toroidal plasma flow, which pushes these unconfined particles outward. All particle sizes considered in this study evaporate within the simulation domain, with boron ions crossing the inner boundary handled by the 1D core impurity model \cite{feng_recent_2014}.

In those first two steps, the study neglects any impurity recycling processes. The spatial distributions of the neutral (B$^{0}$) and ionized (B$^{+1}$ to B$^{+5}$) boron densities are depicted in Figure~\ref{fig:figure2} for mid size B particles ($45$ $\mu$m). In the present scenario, the particles are almost entirely absorbed and ablated near the separatrix at a temperature of $\approx$60 eV. The neutral B atomic sources are peaking where the boron powder particles reach the evaporation temperature. The distribution exhibits elevated concentrations within the SOL for the lower charge states (B$^{+1}$ to B$^{+3}$), characterized by ionization energies spanning $8$ to $38\,\text{eV}$. Conversely, the densities of B$^{+4}$ and B$^{+5}$ ions predominantly peak inside the separatrix. The added boron impurities cause a marginal perturbation in radiative losses, accounting for only 0.1-0.15 MW ($\approx$5-7\% of input power). Radiative cooling effects of boron injection in DIII-D have been reported in \cite{effenberg_2020} and \cite{effenberg_2022}.

A notable increase in boron density is found on the high-field side of the plasma. This inboard peaking of the B density is attributed to the dominant parallel friction force terms associated with the parallel main ion flow, as delineated in equations 2 and 3 and illustrated as Mach flow profile in Figure~\ref{fig:figure2}(h). The parallel main ion flow in the powder deposition region shows a dominant component toward the inboard target (blue region), i.e., pushing the boron impurities effectively poloidally counter-clockwise. Thus, the inward directional bias of the flow culminates in heightened boron densities and fluxes impacting the plasma-facing components on the high-field side.
\begin{figure*}[h!]
\begin{center}
\includegraphics[width=165mm]{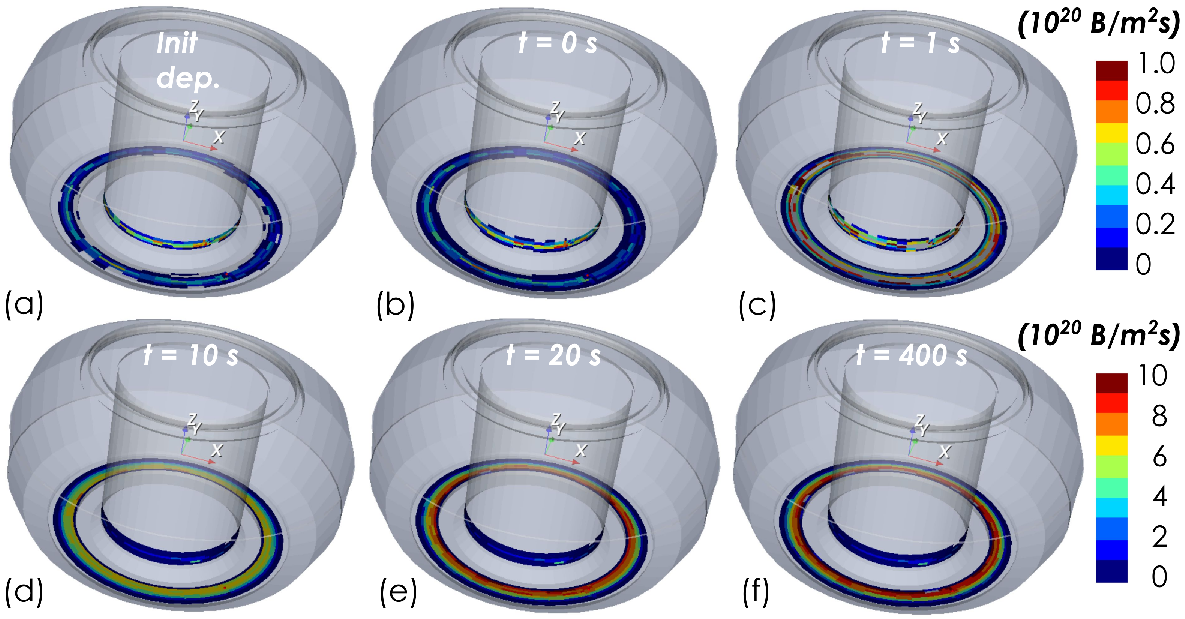}
\caption{\label{fig:figure3} Time evolution of boron flux distribution simulated using WallDYN3D: (a) Initial deposition profile considering first-flight deposition only (excluding reflections, erosion, and redeposition effects); subsequent panels (b-f) illustrate changes at t=0, 1, 10, 20, and 400 seconds, incorporating reflections, erosion, and redeposition dynamics.}
\end{center}
\end{figure*} 
The boron incident flux distributions on the inner and outer divertor targets reveal a substantial asymmetry, with over $95\%$ of the total boron flux directed towards the inboard target according to the EMC3 trace impurity fluid model. The toroidally averaged boron fluxes are shown for all particle sizes in Figure~\ref{fig:figure1}(c). The particle size does not significantly affect this inboard/outboard flux imbalance, with the frictional forces ($\propto V_{fr}\propto M_{||}$) of the main ion flow primarily governing the transport of boron in the plasma scenario considered. However, kinetic modeling in the following will show a reduced imbalance in the fluxes between the inner and outer targets. Moreover, this boron distribution only represents direct deposition from the injection source without considering any subsequent removal or migration upon contact with the target surfaces, i.e., impurities do not recycle. As a result, deposition is narrowly confined to the far SOL, specifically between $Z=-115$ and $-95\,\text{cm}$ on the inboard divertor. Additionally, toroidal asymmetries in the boron deposition on the outer divertor arise from the localized nature of the boron source and the specific orientation of the flux tubes connecting the source region to the divertor targets, as noted in \cite{effenberg_2020}. A limitation of this scenario is that the plasma simulation domain covers only the divertor region, as the plasma grid was originally designed for transport studies rather than plasma-material interaction.

Notably, the substantial in/outboard imbalance in the B fluxes would put this deposition technique into question for low-Z wall coatings and wall conditioning. Contrary to this indication, experimental observations demonstrate substantial boron coatings even within areas that appear to be non-wetted based on the EMC3-EIRENE/DIS calculations  \cite{bortolon_observations_2020, effenberg_hal_2023}). Therefore, the final step in this integrated modeling workflow must consider the effects of subsequent erosion and redeposition processes.
\section{3D boron erosion, migration deposition modeling with WallDYN3D\label{sec:Mixedmaterial}}
The integration of (re)erosion and redeposition processes was achieved by coupling WallDYN3D with the EMC3-EIRENE/DIS plasma solution. For this study, WallDYN3D was expanded to include boron sources (150 $\mu$m) within the plasma domain alongside traditional material targets, enabling the simulation of boron from powder particles.
\begin{figure}[h!]
\begin{center}
\includegraphics[width=85mm]{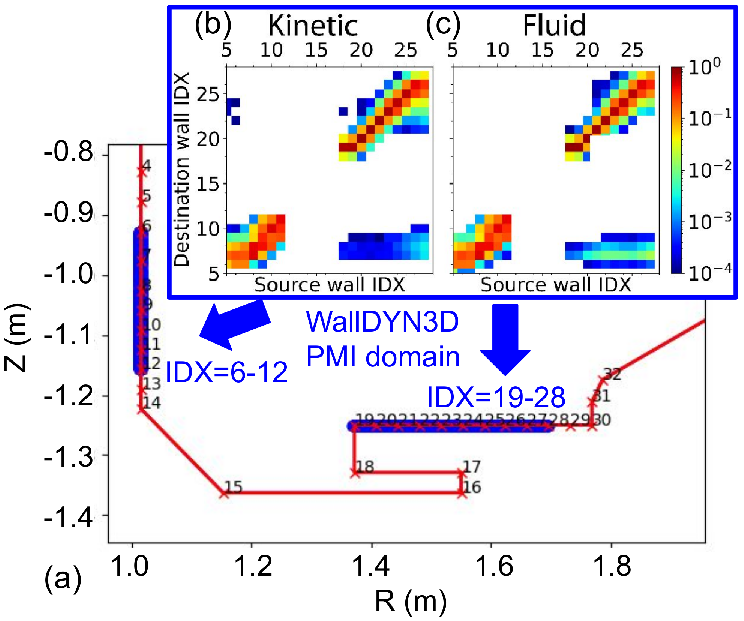}
\caption{\label{fig:figure4} (a) Visualization of the lower divertor region with wall indices (IDX) representing poloidal wall elements. Blue elements are within the simulation domain. Charge state-integrated 2D boron redistribution matrices, as toroidal averages of the 3D matrix, are shown for (b) the fluid model and (c) the kinetic model. "Source wall IDX" indicates where boron impurities are eroded, while "Destination wall IDX" shows where they are likely deposited. Each matrix entry reflects the probability or flux fraction of boron migrating from a specific source to a deposition location.}
\end{center}
\end{figure} 
The source rates were scaled to $10^{20}$ B/s and $10^{21}$ B/s to explore how experimental injection levels of a few milligrams per second up to 20 mg/s may affect the surface distribution and chemical composition (the plasma background was kept constant). Carbon, the predominant intrinsic impurity in the DIII-D tokamak due to its graphite PFCs, was included in the simulation. In the WallDYN3D simulations, the sputtering yields for C, both chemical and charge exchange, were set at 1\% \cite{behrisch_2007}. CX neutral fluxes and average energies are estimated from EMC3-EIRENE background plasma parameters, assuming a Maxwellian-like energy distribution centered on the calculated average energy. Physical sputtering by the background D plasma was also considered, with yields calculated using the energy $3qT_e + 2T_i$ based on the local plasma temperatures. The chemical and charge exchange sputtering for B were neglected. Multiple WallDYN3D time steps were carried out to model the evolution of the areal densities for B and C on the PFCs within the plasma domain. In each step, equations (5) and (\ref{eq:imp_eq6}) are solved, resulting in effective recycling of the impurities.

A 3D visualization of the boron fluxes impacting the divertor targets at an IPD injection rate of $10^{20}$ B/s, as simulated using WallDYN3D, is shown in Figure~\ref{fig:figure3}. Here, the initial boron deposition shows less imbalance between the inner and outer targets compared to the EMC3 fluid model (Figure~\ref{fig:figure1}(c)). In the progress through time steps from 0 to 400 seconds, reflections, erosion, and redeposition are taken into account (Figure~\ref{fig:figure3}(b-f)). After reaching quasi-equilibrium after approximately 20 seconds, there is a significant alteration in the boron flux distribution. The outboard divertor target experiences an increased flux compared to the inboard target, as detailed in Figures \ref{fig:figure3}(e,f).
\begin{figure}[h!]
\begin{center}
\includegraphics[width=85mm]{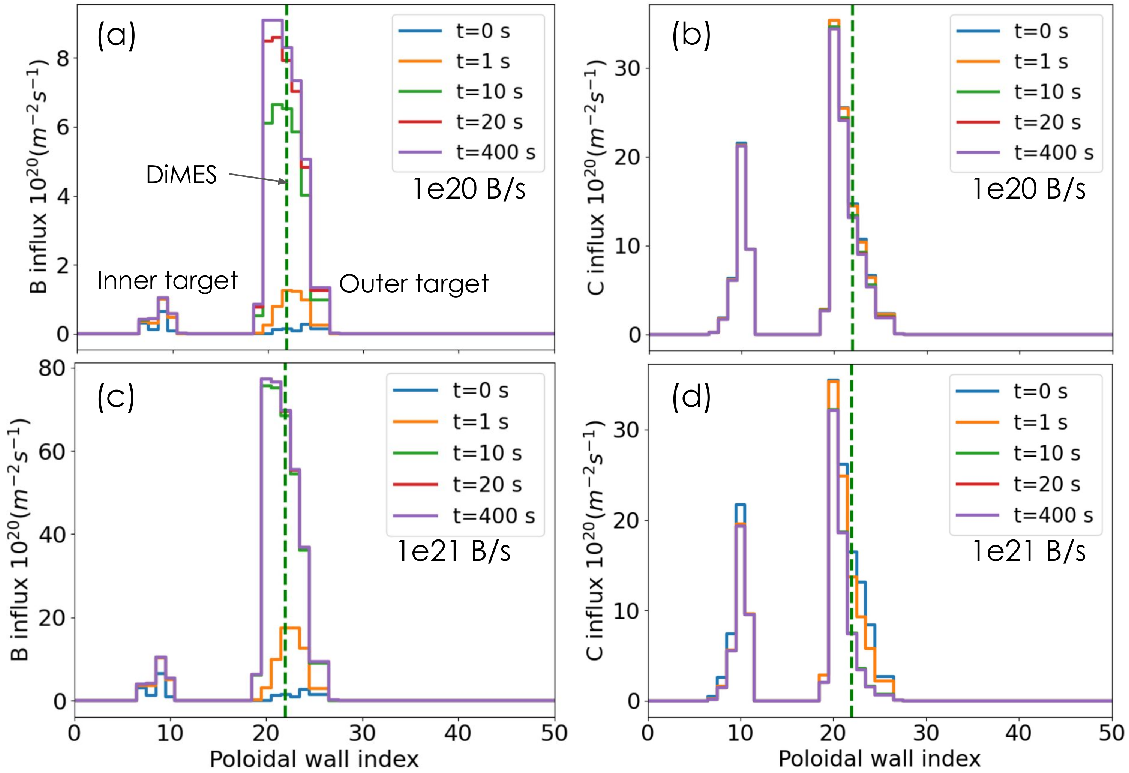}
\caption{\label{fig:figure5} Evolution of WallDYN3D poloidal profiles at the DiMES toroidal location, showing (a) boron and (b) carbon fluxes at a rate of $10^{20}$ B/s, and (c) boron and (d) carbon fluxes at $10^{21}$ B/s for times of 0, 1, 10, 20, and 400 seconds. These profiles account for reflection, erosion, and redeposition. The poloidal position of DiMES is indicated by a dashed line.}
\end{center}
\end{figure}
In WallDYN3D, impurity transport is treated kinetically by default \cite{schmid_2020a}. A comparison between kinetic and fluid modeling was conducted to investigate potential differences in the long-range transport between the high and low-field sides of relatively light B particles. In this case, boron is sourced from the high field side target. The three-dimensional charge-state integrated boron redistribution matrices (see equation (5)) have been toroidally averaged, resulting in two-dimensional representations since this scenario assumed the wall geometry was toroidally symmetric. Figure~\ref{fig:figure4}(a) shows the wall elements and corresponding indices (IDX) in the poloidal plane. The boron redistribution matrices for both the kinetic (Figure~\ref{fig:figure4}(b)) and fluid (Figure~\ref{fig:figure4}(c)) approaches display the probability distribution of where impurities, once eroded, will travel and redeposit, specifically on the divertor wall elements. The diagonal terms indicate the dominating local redeposition of B on the inner ($IDX=6-12$) and outer divertor targets ($IDX=19-28$). The matrices also reveal an absence of transport from the inner to the outer target. However, values in the bottom right sections indicate additional transport from the outer to the inner target. On the inner target, less local redeposition occurs than on the outer target, with less eroded B (and C) returning to their origin (see diagonal elements in Figure~\ref{fig:figure4}(b,c)). This behavior is consistent in both the fluid and kinetic treatment qualitatively and quantitatively within the WallDYN3D model. The remaining differences between the kinetic and fluid models stem from how they handle impurity transport near the boron injection location. The fluid model assumes immediate equilibrium with the plasma, simplifying the process and leading to overestimated long-range transport. In contrast, the kinetic model accounts for the time needed for boron ions to accelerate and thermalize, resulting in more accurate, localized deposition.

It is important to note that the simulation domain encompasses only a significant portion of the divertor targets within the DIII-D tokamak rather than the entire wall structure. Simulation results indicate that approximately 30\% of the boron flux is not accounted for within this domain, suggesting that it would be deposited on wall elements outside the plasma simulation domain.
\section{Impact of boron source strength on mixed-material coating composition\label{sec:BCratio}}
The dynamics of the B-C surface composition are affected by erosion and reflection properties on wall elements. One-dimensional boron and carbon flux profiles at the toroidal DiMES location (dashed vertical line, \cite{rudakov_2017}) are shown in Figure~\ref{fig:figure5} for time steps of 0, 1, 10, 20, and 400 seconds, with boron influx rates of $10^{20}$ and $10^{21}$ B/s (see poloidal wall index in Figure \ref{fig:figure4}(a)). The WallDYN3D calculations indicate that boron flux increases significantly due to reerosion, migration, and redeposition while carbon flux decreases. At the lower boron rate, boron flux to the high-field side target (inner peak) increases over time from $\approx6 \times 10^{19}$ to $1.2 \times 10^{20}$ m$^{-2}$s$^{-1}$, and more dramatically on the outboard side (outer peak) from $\approx 3 \times 10^{19}$ to $9 \times 10^{20}$ m$^{-2}$s$^{-1}$. The carbon flux sees only a slight reduction. Conversely, at the higher boron rate ($10^{21}$ B/s), peak levels stabilize quickly, within 1 second at the inboard target, with carbon levels dropping by about 10\%. 
\begin{figure}[h!]
\begin{center}
\includegraphics[width=85mm]{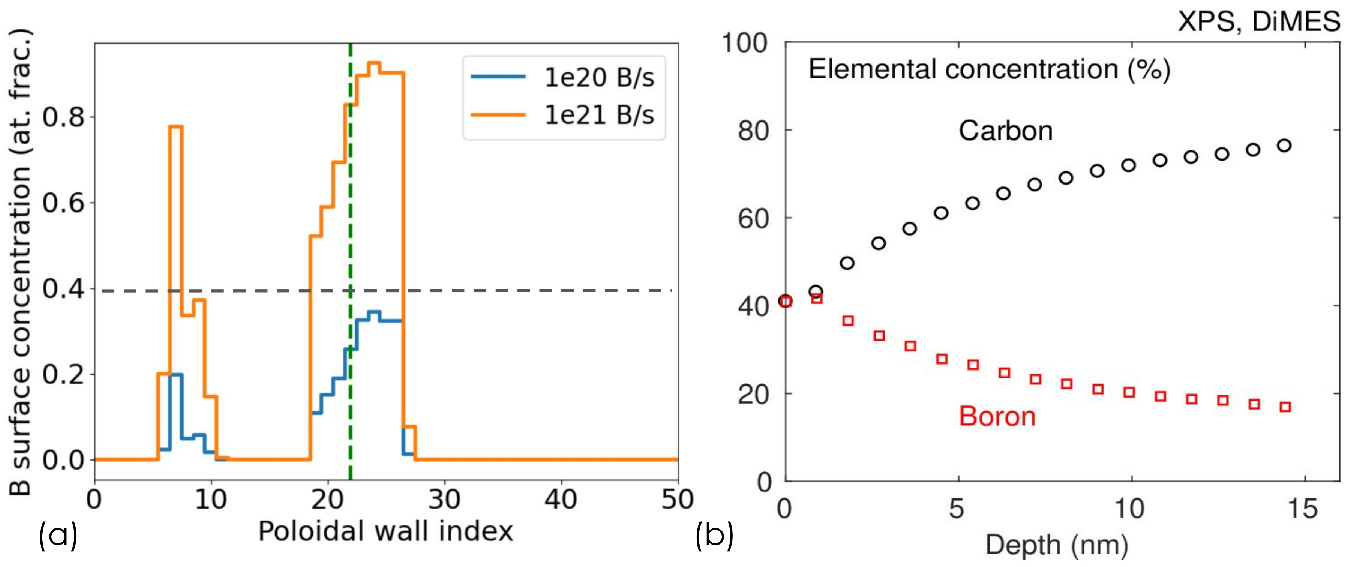}
\caption{\label{fig:figure6} (a) Boron surface concentrations after 400 iterations for moderate ($10^{20}$ B/s) and high ($10^{21}$ B/s) boron source strengths at the toroidal DiMES location. The horizontal broken line indicates the experimental boron concentration on DiMES. (b) XPS depth profile from a similar experiment shows 40\% boron (40\% carbon, with 20\% oxygen and other impurities not shown) \cite{bortolon_observations_2020}.}
\end{center}
\end{figure}
Most of the eroded boron (B) and carbon (C) recycle in front of the target through repeated erosion and re-deposition. Local re-deposition is lower on the inner target than the outer target, as shown by the migration matrix. Consequently, less of the eroded material returns to its origin on the inner target, whereas the outer target retains more due to higher re-deposition rates. The erosion yields of B layers are higher than those of carbon, and for a 50:50 B-to-C surface, the erosion flux is greater for B. The increased B influx is primarily driven by local recycling, not long-range transport (as shown in the previous section). The influx of B on the outer target is higher simply because more erosion and reflection flux is coming back to the target. The surface composition is due to the dynamic balance of erosion vs. influx. On the outer target, high re-deposition rates result in approximately zero net erosion of C and minor accumulation of B, supported by the continuous B source from the impurity powder dropper. In contrast, the inner target exhibits net erosion of C and nearly zero net deposition of B due to lower re-deposition. As stated before, some B is lost (especially from the inner target).

Adjusting the boron injection rate modifies the surface composition of divertor targets, as shown in Figure~\ref{fig:figure6}. With a boron influx rate of $10^{20}$ B/s, the B surface concentrations on the inner and outer targets reach approximately 35\%, with boron area densities ranging from $0.3-1.4\times10^{16}$ cm$^{-2}$. Increasing the influx rate tenfold to $10^{21}$  B/s significantly enriches the boron on the target surfaces, potentially forming nearly pure boron layers, with area densities between $1.8$ and $3.6\times10^{16}$ cm$^{-2}$. This saturation phenomenon contrasts with lower B injection rates, where the surface composition sustains a mixed B-C coating, approximating a B concentration closer to experimental findings, with B and C concentrations each being 40\% as shown in Figure \ref{fig:figure6}(b). This effect can be attributed to the differential sputtering rates of B and C, which minimize the overall sputtering yields. It has been observed in previous DIII-D IPD experiments and traditional boronization procedures in tokamaks like TEXTOR and ASDEX-Upgrade \cite{bortolon_observations_2020, winter_1989, asdex_1990}. For IPD experiments, however, achieving a balanced B:C concentration typically required B powder rates of 10-20 mg/s or $10^{21}$ B/s. Discussions in the literature, such as in \cite{bortolon_observations_2020}, highlight a marginal dependence of surface composition on boron influx rates, underscoring the challenges of attaining pure boron layers in environments with significant carbon presence. Differences between modeling and experiment may be partially explained by the simulation domain focusing on the divertor region, with the main chamber wall not yet included. Carbon wall elements beyond the divertor might contribute to the total C influxes due to additional plasma-wall interactions (e.g., main wall CX-erosion) not considered in the present study.
\section{Summary and conclusions\label{sec:Conclusion}}
A new comprehensive modeling approach for real-time wall conditioning and in-situ boron powder coating is presented, utilizing three advanced models: EMC3-EIRENE, DIS, and WallDYN3D \cite{feng_recent_2014, nespoli_2021, schmid_2011, schmid_2018, schmid_2020}. These models collectively address 3D plasma transport, powder particle migration and ablation, and mixed-material surface evolution.
Ablation modeling with EMC3-EIRENE/DIS shows that boron particles of up to 150 $\mu$m evaporate at the separatrix. The EMC3 trace impurity fluid modeling shows that initial B fluxes are higher on the inboard divertor target due to the directional frictional forces from the main ion flow in the evaporation region. This effect aligns with findings from EMC3-EIRENE simulations using a single atomic B source in the same plasma scenario previously \cite{effenberg_2020}. This trend is consistent with beryllium transport modeling for JET and ITER using ERO2.0 \cite{romazanov_2024}. However, first-flight deposition based on EMC3-EIRENE/DIS alone neglects impurity recycling, resulting in significant asymmetries in the divertor B fluxes concentrated in a narrow, far SOL region. The powder particle mass was not found to affect the distribution of B fluxes between the inboard and outboard divertor targets for particle sizes 5-250 $\mu$m in the present plasma scenario.

Additional mixed-material sources and sinks, i.e., erosion, reflection, and redeposition, were taken into account using the WallDYN3D code. The code was extended to include external impurity sources from powder injection. The kinetic treatment of impurities in WallDYN3D initially shows a reduced imbalance in B fluxes between the inner and outer divertor targets compared to the EMC3 fluid model. B and C primarily recycle locally through erosion and redeposition, with the outer target showing a higher B influx due to greater erosion and reflection flux. This leads to near-zero net erosion of C and minor B accumulation on the outer target, supported by local recycling and the continuous B source, while the inner target shows net erosion of C and less local B redeposition. The surface composition is governed by the dynamic balance of erosion and redeposition rather than long-range transport. The approach also replicates B surface concentrations on the outer divertor close to experimental findings. However, the model underpredicts the B fluxes necessary to achieve this balance, likely partially due to a limited plasma grid domain that mainly covers the divertor region and neglects potential additional C sources from the remaining wall. Therefore, extending the simulation grid for the whole wall will be considered crucial for future work.

This integrated modeling approach provides a new framework for studying real-time wall conditioning and in-situ coating in advanced tokamak scenarios. The results suggest that poloidal and toroidal coverage of divertor plasma-facing components with boron can be achieved through single-point B powder injection within the limitations of the model and its implementation. The insights garnered from this work will inform similar studies planned for ITER \cite{snipes_2024}.
\section*{Acknowledgements}
This material is based upon work supported by the U.S. Department of Energy, Office of Science, Office of Fusion Energy Sciences, using the DIII-D National Fusion Facility, a DOE Office of Science user facility, under Awards DE-AC02-09CH11466, DE-FC02-04ER54698, and DE-AC05-00OR22725.
The United States Government retains a non-exclusive, paid-up, irrevocable, world-wide license to publish or reproduce the published form of this manuscript, or allow others to do so, for United States Government purposes. The data that support the findings of this study are openly available in Zenodo at https://zenodo.org/records/14176757.
\section*{Disclaimer}
This report was prepared as an account of work sponsored by an agency of the United States Government. Neither the United States Government nor any agency thereof, nor any of their employees, makes any warranty, express or implied, or assumes any legal liability or responsibility for the accuracy, completeness, or usefulness of any information, apparatus, product, or process disclosed, or represents that its use would not infringe privately owned rights. Reference herein to any specific commercial product, process, or service by trade name, trademark, manufacturer, or otherwise does not necessarily constitute or imply its endorsement, recommendation, or favoring by the United States Government or any agency thereof. The views and opinions of authors expressed herein do not necessarily state or reflect those of the United States Government or any agency thereof.

\bibliography{mainbib}
\end{document}